\documentclass[]{spie}

\usepackage{amsmath,amsfonts,amssymb}
\usepackage{longtable}
\usepackage{graphicx}
\usepackage[acronym]{glossaries}
\usepackage{hyperref}
\usepackage{cleveref}
\usepackage{caption}
\usepackage{subcaption}
\usepackage{siunitx}

\usepackage{enumitem}
\setlist{nosep}

\usepackage{tikz}
\usetikzlibrary{shapes,arrows,dsp,fit,positioning}
\usepackage[americanresistors,americaninductors]{circuitikz}

\title{Development of TRL5 Firmware for Tuning, Biasing, and Readout of Kilopixel TES Bolometer Arrays}

\author[a]{Graeme Smecher}
\author[b]{Jean-François Cliche}
\author[c]{Matt Dobbs}
\author[d]{Joshua Montgomery}
\affil[a]{Three-Speed Logic, Victoria, Canada}
\affil[b,c,d]{McGill University, Montreal, Canada}

\authorinfo{E-mail: gsmecher@threespeedlogic.com}

\newacronym[\glslongpluralkey={Digitizer Assemblies}]{da}{DA}{Digitizer Assembly}
\newacronym[\glslongpluralkey={SQUID Controller Assemblies}]{sca}{SCA}{SQUID Controller Assembly}
\newacronym[\glslongpluralkey={Signal Processing Assemblies}]{spa}{SPA}{Signal Processing Assembly}
\newacronym{adc}{ADC}{Analog-to-Digital Converter}
\newacronym{bram}{BRAM}{Block Random Access Memory}
\newacronym{cdr}{CDR}{Clock/Data Recovery}
\newacronym{cic}{CIC}{Cascaded Integrator-Comb}
\newacronym{clb}{CLB}{Combinatorial Logic Block}
\newacronym{dac}{DAC}{Digital-to-Analog Converter}
\newacronym{dan}{DAN}{Digital Active Nulling}
\newacronym{dds}{DDS}{Direct Digital Synthesizer}
\newacronym{dpu}{DPU}{Data Processing Unit}
\newacronym{dsp}{DSP}{Digital Signal Processing}
\newacronym{ecc}{ECC}{Error-Correcting Code}
\newacronym{fdir}{FDIR}{Fault Detection, Isolation, and Recovery}
\newacronym{fdm}{FDM}{Frequency-Domain Multiplexing}
\newacronym{fft}{FFT}{Fast Fourier Transform}
\newacronym{fir}{FIR}{Finite Impulse Response}
\newacronym{fpga}{FPGA}{Field-Programmable Gate Array}
\newacronym{icd}{ICD}{Interface Control Document}
\newacronym{iic}{IIC}{Inter-Integrated Circuit}
\newacronym{iob}{IOB}{Input/Output Block}
\newacronym{ird}{IRD}{Interface Requirements Document}
\newacronym{jaxa}{JAXA}{Japanese Aerospace Exploration Agency}
\newacronym{lut}{LUT}{Look-Up Table}
\newacronym{lvds}{LVDS}{Low-Voltage Differential Signaling}
\newacronym[\glslongpluralkey={Power Controller Assemblies}]{pca}{PCA}{Power Controller Assembly}
\newacronym{pfb}{PFB}{Polyphase Filter Bank}
\newacronym{pll}{PLL}{Phase-Locked Loop}
\newacronym{rll}{RLL}{Resistance-Locked Loop}
\newacronym{scu}{SCU}{SQUID Controller Unit}
\newacronym{seu}{SEU}{Single-Event Upset}
\newacronym{spi}{SPI}{Serial Peripheral Interface}
\newacronym{spt}{SPT}{South Pole Telescope}
\newacronym{spu}{SPU}{Signal Processing Unit}
\newacronym{squid}{SQUID}{Superconducting Quantum Interference Device}
\newacronym[\glsshortpluralkey={TESes}]{tes}{TES}{Transition Edge Sensor}
\newacronym{tmr}{TMR}{Triple Modular Redundancy}

\begin{document} 
\maketitle

\begin{abstract}
	The next generation of space-based mm-wave telescopes, such as \acrshort{jaxa}'s LiteBIRD mission, require focal planes with thousands of detectors in order to achieve their science goals.
	Digital frequency-domain multiplexing (dfmux) techniques allow detector counts to scale without a linear growth in wire harnessing, sub-Kelvin refrigerator loads, and other scaling problems.
	In this paper, we describe the Digital Signal Processing (DSP) firmware executed in the design's \acrfullpl{fpga}.
	This firmware is responsible for synthesizing bias tones, performing dynamic feedback control of the bolometer voltage bias and/or Superconducting Quantum Interference Device (SQUID) nuller currents, demodulating and decimating bolometer channels into science data, and streaming the results for storage and eventual downlink.
	We describe how this firmware has been tailored for LiteBIRD, including the control path, improvements to power- and resource-efficiency, the addition of radiation-mitigation functions, and the integration of new bolometer biasing schemes that may help mitigate mission-specific design challenges.
	This paper is a companion piece to the description of the electronics platform in which the firmware operates.
\end{abstract}

\keywords{Bolometer, FPGA, CMB, satellite, SQUID}

\section{INTRODUCTION}
\label{sec:intro}

Over the past decade, superconducting detectors in mm-wave astronomy have matured from experimental development to proven technology.
This maturation has been driven by a fundamental need to scale: because it is possible to build detectors that are photon-noise limited, the ability to advance science using them is chiefly enabled by successive experiments with larger focal planes and increasing numbers of detectors.

For example: the receiver for the original \acrfull{spt} in Antarctica was commissioned in 2007 with 960 bolometers \cite{Schaffer2011}.
The follow-on SPTPol experiment was commissioned in 2011 with 1,536 bolometers \cite{Austermann2012}.
The SPT-3G camera was commissioned in 2017 with 16,000 bolometers \cite{Sobrin2022}.
SPT-3G is expected to be replaced with the SPT-Slim pathfinder and SPT-3G+ camera in the coming years, and will have a detector count on the order of 35,000 \cite{Dibert2021}.
This growth in detector counts is matched by other experiments across the sector, including the CMB-S4 telescope which is expected to have 500,000 detectors when it is commissioned \cite{Abazajian2016}.

In this paper, we present state-of-the-art readout electronics for a satellite-borne mm-wave telescope such as the LiteBIRD experiment\cite{LiteBIRD2022}.
This readout supports frequency-domain multiplexed \acrfull{tes} bolometers, an evolving technology stack known as dfmux \cite{Montgomery2022, Dobbs2007}.
The signal paths described here achieve multiplexing densities of 128 bolometers per SQUID, and 16 \acrshortpl{squid} per \acrshort{fpga} using a commercially available, space-qualified \acrshort{fpga}.
This paper is a companion piece to Ref.~\citenum{Smecher2022a}, which describes the hardware platform and geometry in detail.




We proceed as follows:
\begin{itemize}
	\item We briefly review the fundamental operation of a TES bolometer readout using dfmux techniques;
	\item We show a structural overview of the system's firmware;
	\item We provide an overview of the DSP algorithms implemented within each block;
	\item We contrast this design with previously published information on dfmux implementations;
	\item We describe the performance of the system; and
	\item We conclude with a summary of the firmware's current status and next steps.
\end{itemize}

\section{DIGITAL FREQUENCY-DOMAIN MULTIPLEXING (DFMUX)}

Digital frequency-domain multiplexing (dfmux) is shorthand for a combination of technologies used for TES bolometer readout \cite{Montgomery2022, Dobbs2007}.
The following is a simplified description, intended to motivate the digital signal processing firmware described in the material that follows.

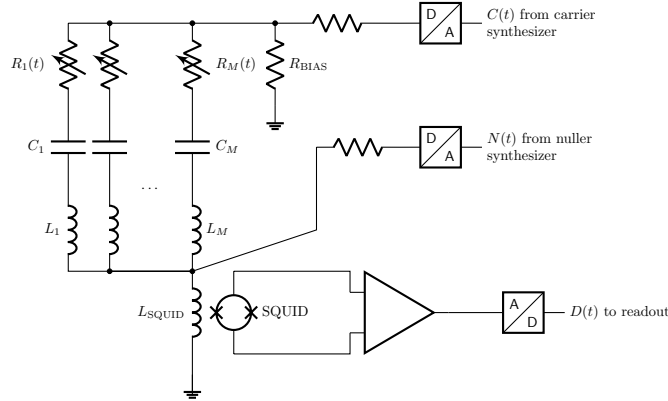
\begin{figure}
	\centering
	\begin{circuitikz}[scale=0.55, transform shape]
	\draw
		(1,9) -- (6,9) to [R] (9,9) to [dac] (11,9)

		(1,9) to [vR, -, l_=$R_1(t)$] (1,7)
		to [C, l_=$C_1$] (1,5)
		to [L, l_=$L_1$] (1,3) -- (2,3)
		-- (3,3) -- (4,3)

		(2,9) to [vR, *-] (2,7)
		to [C] (2,5)
		to [L, -*] (2,3) -- (4,3)

		(3,5) node[] {$\cdots$}

		(4,9) to [vR, *-, l=$R_M(t)$] (4,7)
		to [C, l=$C_M$] (4,5)
		to [L, l=$L_M$, -*] (4,3)

		(6,7) node[ground] {}
		to [R, -*, l_=$R_\mathrm{BIAS}$] (6,9)

		(4,3) to [L, l_=$L_\mathrm{SQUID}$] (4,1)
		to node[ground] {} (4,0)

		(7,6) to [R] (9,6) to [dac] (11,6)
		(7,6) |- (7,4) -- (4,3)

		(9,2) node[plain amp] (amp) {}

		(amp.-) |- (5,3)
		(amp.+) |- (5,1)

		(5,3) to [squid,l={~SQUID}] (5,1)

		(amp.out) -- (11,2) to [adc] (13,2)
		;

		\node[right, text width=3cm] at (11,9) {$C(t)$ from carrier synthesizer};
		\node[right, text width=3cm] at (11,6) {$N(t)$ from nuller synthesizer};
		\node[right, text width=3.5cm] at (13,2) {$D(t)$ to readout};

	\end{circuitikz}
	\caption{Simplified schematic of the cryogenic portion of a dfmux system.}
	\label{fig:cold-schematic}
\end{figure}

\Cref{fig:cold-schematic} shows a simplified cryogenic schematic associated with a dfmux system.
Our goal is to create a system that measures the current through each bolometer in response to an applied bias voltage.
Because each bolometer behaves as a time-varying resistance $R_i(t)$ dependent on incident optical power, the resulting current measurement faithfully encodes what the bolometer ``sees''.
To allow bolometers to be measured individually, each is paired with a distinct LC filter (typically between 1-6 MHz; see e.g. Ref.~\citenum{Montgomery2022}) that assigns it a distinct frequency.

A bias voltage is established across $R_\mathrm{BIAS}$ as follows:
The readout electronics (not shown in \Cref{fig:cold-schematic}) synthesize a sum of sinusoids (the ``carriers''), each with programmable frequency, phase, and amplitude.
Because the impedance of $R_\mathrm{BIAS}$ is small compared to the LCR networks and $L_\mathrm{SQUID}$ at all carrier frequencies, the bias resistor is the dominant path for carrier current.
This arrangement establishes a programmable bias voltage across $R_\mathrm{BIAS}$ that is insensitive to the bolometer branches of the circuit and proportional to the carrier current.

In order to measure each bolometer's resistance, we wish to extend this programmable voltage bias from $R_\mathrm{BIAS}$ to each bolometer in the neighbourhood of its LC filter's resonant frequency and measure the resulting current.
To minimize the voltage developed across $L_\mathrm{SQUID}$, we inject a second sum of sinusoids (the ``nullers'') intended to cancel the carrier currents passing through each bolometer and hence through the \acrshort{squid} inductor.
To the extent that these nuller and carrier currents cancel each other, no current passes through $L_\mathrm{SQUID}$ and a virtual ground is created at the summing junction.
This virtual ground also linearizes the \acrshort{squid} transimpedance amplifier, and removes crosstalk between bolometers due to the common impedance.\cite{Dobbs2012, Montgomery2022}
This technique is known as \acrfull{dan} \cite{deHaan2012}.

Any residual bolometer currents which are not nulled pass through the \acrshort{squid} inductor, which combines with the \acrshort{squid} to form a transimpedance amplifier.
These residuals are used to update the nuller parameters in real time to maintain the virtual ground at $L_\mathrm{SQUID}$.

Science data consists of estimates of current through each bolometer over time.
These estimates are calculated from two data sources:
\begin{itemize}
	\item
		Bolometer voltages are known a-priori and are a function of carrier parameters (the statically programmed frequency, phase, and amplitude settings used to synthesize the carrier comb.)
		These parameters depend on as-built characteristics of the focal plane (e.g. the locations of bolometer resonances), and are determined during commissioning and programmed into the signal path as part of regular tuning operations; and
	\item
		Bolometer currents are, within the bandwidth of the nuller feedback loop, equal and opposite to nuller currents.
		These currents are generated dynamically via the nuller feedback loop, and reconstructing them numerically requires knowledge of both a-priori parameters (i.e. frequency, phase, digital loop gain) and the continuously updated settings that form the control loop (i.e. complex amplitudes).
		Complex amplitudes for each active bolometer channel are decimated, timestamped, packetized, and shipped off-board for storage, downlink, and eventually analysis.
\end{itemize}

This description neglects crosstalk and parasitic losses in the system that weaken voltage bias or decouple the nuller and bolometer currents.
A treatment of these effects and techniques to mitigate them are given in Ref.~\citenum{Montgomery2022}.

\section{STRUCTURAL OVERVIEW}

We begin with a brief description of the hardware context surrounding the firmware.
We then provide a structural overview of the firmware itself.

\subsection{Design Context}

The firmware described here operates on an \acrshort{fpga} within the \acrfull{spu}, which performs digital signal-processing operations associated with a satellite-borne dfmux readout.
Each \acrshort{spu} supports up to 16 \acrshort{squid} channels, with up to 128 TES bolometers per \acrshort{squid}.
A given satellite may integrate a number of \acrshortpl{spu} in order to increase the total bolometer count and provide for redundancy.
(For the LiteBIRD mission, there are 6 primary \acrshortpl{spu} and 6 cold-spare \acrshortpl{spu}.)

Within each \acrshort{spu} are several sub-assemblies:
\begin{itemize}
	\item A \acrfull{spa}, which houses an \acrshort{fpga} and associated electronics (firmware for this \acrshort{fpga} is the primary subject of this work),
	\item Up to 4 \acrfullpl{da} that contain \acrshortpl{adc}, \acrshortpl{dac}, and associated analog signal-processing electronics, and
	\item A \acrfull{pca} that provides an isolated power supply at voltages appropriate for the \acrshort{spu}.
\end{itemize}

Each \acrshort{da} within the \acrshort{spu} interfaces with an \acrfull{sca}, which resides within an \acrfull{scu} and contains the last stage of analog signal processing before the cryogenic hardware on focal plane itself.
The \acrshort{spu} powers and controls the \acrshort{scu}.

Each \acrshort{spu} interfaces with a primary and a redundant \acrfull{dpu}, which is the only interface between the \acrshort{spu} and spacecraft bus.
All commands, data, and clocking from the spacecraft bus to the \acrshort{spu} occur over these links.
Only one link (primary or redundant) may be active at any time.

A complete system context diagram showing module definitions, multiplicities, and interconnections is given in Ref.~\citenum{Smecher2022a}.

\subsection{Firmware Overview}

\begin{figure}
	\centering
	\includegraphics{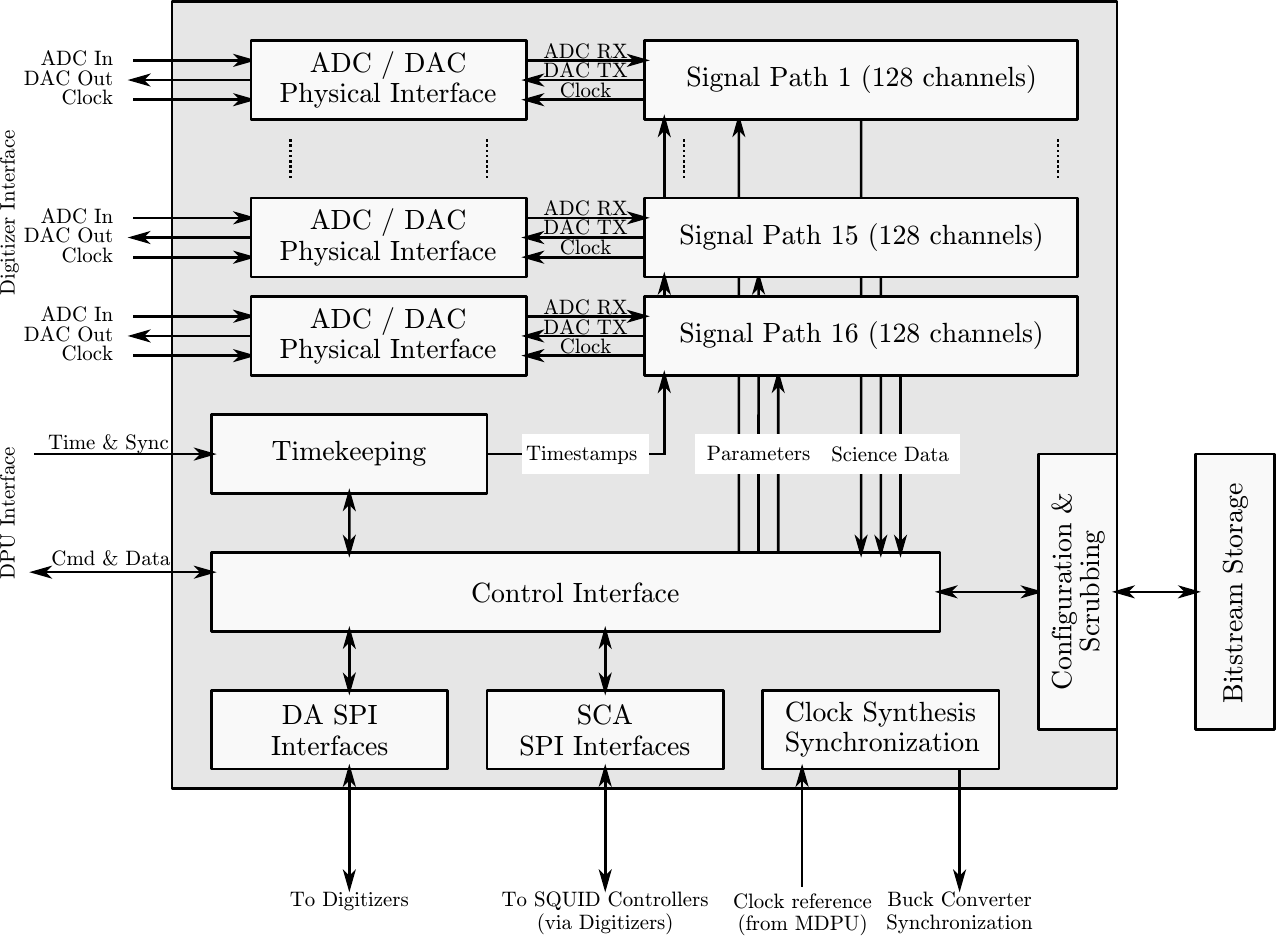}
	\vspace{2ex}
	\caption{
		Top-level block diagram of SPA firmware.
		This firmware interfaces with 16 \acrshortpl{squid}, each requiring 2 DAC channels (carrier/nuller) and 1 ADC channel at 20 MSPS each.
		The design consists of 16 signal paths (which dominates resource usage), and a number of relatively small housekeeping and interface blocks.
	}
	\label{fig:top-level-firmware}
\end{figure}

A structural diagram of \acrshort{spa} firmware is shown in \Cref{fig:top-level-firmware}.
Firmware consists of the following major elements:
\begin{itemize}
	\item
		Signal paths, which implement the synthesis, demodulation, and readout algorithms central to the readout;
	\item
		Spacecraft control interfaces that manage command and data communications with the mission \acrfull{dpu};
	\item
		\acrshort{da} and \acrshort{sca} control interfaces, managing communications with connected assemblies;
	\item
		Scrubbing and bitstream management logic associated with ensuring reliable, radiation-hardened \acrshort{fpga} operation; and
	\item
		Clock synthesis, reset, and synchronization modules associated with generating and distributing clean, globally synchronized clocks to elements within the \acrshort{spa} and connected assemblies.
\end{itemize}

With the exception of scrubbing and bitstream management, there is no software in the system.
``Firmware'' is used as shorthand for HDL code (VHDL and/or SystemVerilog).

To give a sense of the relative scale and complexity of the units shown in \Cref{fig:top-level-firmware}, an \acrshort{fpga} floorplan is shown later in \Cref{fig:top-level-floorplan}.
The overwhelming majority of \acrshort{fpga} resources are either used by signal paths, or unoccupied.
Other blocks in \Cref{fig:top-level-firmware} are minor in comparison.

Interfaces described here typically have some multiplicity associated with them (to accommodate mission geometries and cross-strapping).
Because we focus here on logic design, we do not always specify multiplicities.
A description of overall unit and mission geometries, as well as redundancies and cross-strapping, is provided in the hardware companion paper\cite{Smecher2022a}.

\subsubsection{Spacecraft I/O Interfaces}
\label{sec:spacecraft-io}

Each spacecraft I/O interface is an 8b/10b encoded\cite{Widmer1983}, full-duplex serial link at 50 Mbps.
These links carry commands and responses, as well as streamed science data and housekeeping telemetry.
They are DC-coupled, and each is accompanied by a source-synchronous 25 MHz clock in both TX and RX directions.
Although 8b/10b encoding is not strictly necessary with a DC-coupled, source-synchronous link, the comma symbols available in this scheme provide bit/frame alignment and idle-state detection.

All data on spacecraft I/O links is encapsulated in packets called ``enclosures'', which provide a consistent description of the packets' type, size, and sequence information.
Enclosures allows packet data to be organized, forwarded, and filtered by generic packet-processing firmware that does not deeply understand its content.

All commands from the \acrshort{dpu} (for example, focal plane tuning operations) are initiated by the \acrshort{dpu} and are expressed as reads or writes into the SPA's register space.
Each of these read/write request packets generates a corresponding response packet confirming the read or write completed and including results, if any.
Because link and signal-path latency is expected to limit overall tuning performance, sequences of independent read/write commands will be issued in bulk rather than requiring each request to generate a response before subsequent requests are issued.
Although it requires tuning algorithms to be expressed and coded with parallelism in mind, this asynchronous I/O style is critical to limit time spent tuning the focal plane.

The SPU transmits science data autonomously as it emerges from the datapath.
No flow control (e.g. acknowledgements or retransmissions) or error-correction codes are used in the science-data stream.

\subsubsection{Configuration and Scrubbing}

On power-up, the SPU's \acrshort{fpga} loads a bitstream from on-board \acrshort{spi} flash.
This bitstream allows the \acrshort{spu} to communicate with the \acrshort{dpu} and is not in-flight upgradeable.
The \acrshort{dpu} then instructs the \acrshort{spu} to load a complete bitstream, which is in-flight upgradeable.
On-board flash storage is large enough for two such bitstreams to be maintained, allowing in-system upgrades to proceed while keeping a ``fall-back'' bitstream in case the upgrade fails.

An L2 mission such as LiteBIRD must contend with a radiation environment that is mostly benign, and occasionally extremely active.\cite{Barth1999}
The \acrshort{fpga} is susceptible to \acrfullpl{seu}, which flip bits in the \acrshort{fpga}'s configuration memory (that is, the \acrshort{fpga}'s description of the synthesized RTL) or design registers (the contents of flip-flops or other state within that design.)
Firmware includes an \acrshort{seu} scrubber that detects and corrects radiation-induced upsets in portions of the \acrshort{fpga}'s configuration bitstream.
This scrubber provides partial \acrshort{seu} protection without the resource cost of a \acrfull{tmr} approach.
Bit flips in user state are managed through a variety of design interventions, including \acrfullpl{ecc} and periodic self-reset.
The flight scrubber will be completed during a future design phase.
Preliminary test results for \acrshort{seu} susceptibility are summarized in \Cref{sec:conclusions-seu}.

\subsubsection{Subassembly I/O Interfaces}

The low-speed interfaces for connected assemblies (SCA, DA) are either single-bit I/Os or simple variations on \acrshort{spi} or \acrshort{iic}.
These interfaces control data converters (DACs, ADCs) and control signals associated with SQUID and bolometer biases. These I/O interfaces are simple and are not described in greater detail here.

\subsubsection{Clock Synthesis, Timekeeping, and Synchronization}

The \acrshort{spu} operates on either a primary or redundant 10 MHz clock supplied via the primary or redundant \acrshort{dpu} interfaces.
These clocks are routed through \acrshort{lvds} receivers directly to the \acrshort{fpga} within each \acrshort{spa}.
From the \acrshort{fpga} perimeter, both of these clocks are routed directly routed to a \acrshort{pll} to provide synthesized clocks for the remainder of the system.

Alongside the master clock, \acrshortpl{dpu} distribute a synchronization strobe.
This ``sync'' signal allows deterministic distribution  of synchronization events across multiple \acrshortpl{spu}.
Synchronization events are used as follows:

\begin{itemize}
	\item To ensure timestamps across multiple \acrshortpl{spu} are consistent, and
	\item To ensure synchronization of signal-path element state (e.g. decimators) within and between \acrshortpl{spu}.
\end{itemize}

Timestamps in the \acrshort{spa} are simple 48-bit counters incremented using the 10 MHz ``master'' clock.
The values in these counters may be synchronized to a spacecraft-wide value under \acrshort{dpu} control by pre-programming this value into FPGA registers within the \acrshort{spa} and triggering a ``sync'' event to latch them.
Because timekeeping across \acrshortpl{spu} shares this common clock, synchronization is part of \acrshort{spu} power-up operations and need not be repeated.

Synchronization of signal-path elements (in order to ensure matching timestamps across science data packets from different \acrshortpl{spu}) occurs as follows.
Signal-path elements are placed in reset by timekeeping logic within each \acrshort{spu}.
A global ``sync'' event brings the datapaths out of reset on the same edge of the active master clock.

\section{Signal Path}

The bulk of the firmware design effort (and correspondingly, the bulk of occupied \acrshort{fpga} fabric) is dedicated to the system's real-time signal path.
In the following sections, we describe the design and implementation of a single SQUID channel's signal path.
As shown in \Cref{fig:top-level-firmware}, there are 16 such signal paths per SPA.
Each of these 16 signal paths function independently.

\subsection{Overview}
\label{conceptual-overview}

\begin{figure}
	\includegraphics[width=\textwidth]{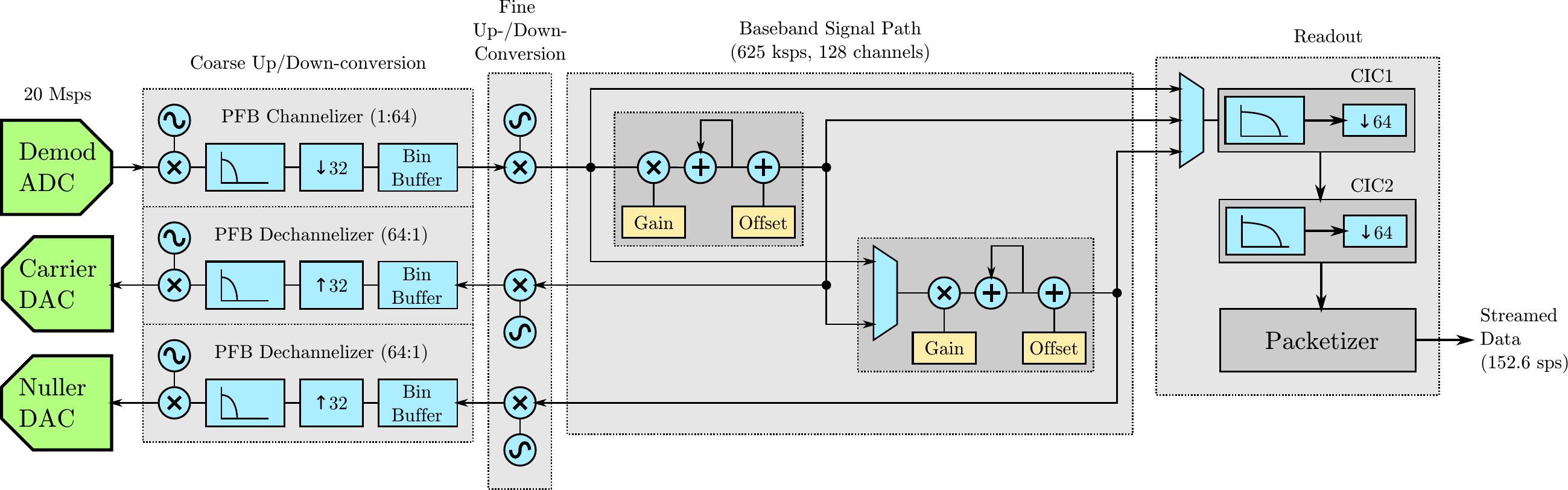}
	\caption{
		Conceptual view of the dfmux signal path.
		This diagram shows 1 of 16 SQUID channels, each of which support up to 128 bolometers.
	}
	\label{fig:dan-path}
\end{figure}

A conceptual model of the readout signal path is shown in \Cref{fig:dan-path}.
This image shows the downconversion, feedback loop, and upconversion processes seen by each bolometer channel.
It also shows the parameter settings (frequencies, phases, and amplitudes) used within the datapath and configured by the \acrshort{dpu}.
This signal path services the four core functions of the dfmux readout system:

\begin{itemize}
\item
  Ingestion and downconversion of science data to multiple baseband channels,
\item
  Use of demodulated science data to generate time-varying nuller or carrier parameters,
\item
  Synthesis of nuller and carrier sinusoids from fixed or time-varying parameters, and
\item
  Decimation and encapsulation of readout signals for storage and downlink.
\end{itemize}

\Cref{fig:dan-path} shows a single datapath, which is an accurate reflection of how it is constructed.
In order to scale up to 128 bolometers per SQUID, this datapath is time-multiplexed.
It operates using an internal clock rate of 200 MHz, allowing it adequate time to compute 128 channels at 625 ksps by rapidly processing each bolometer channel in sequence.

In the following sections, we describe the structure of selected portions of the signal path in detail.

\subsection{Coarse Up-/Down-conversion}

Because the up-conversion process is a dual of the downconversion process, we focus on downconversion here and assume a matched geometry on the synthesis side of the system.

First, the 20 MSPS signal streams received from the system's ADCs are channelized into 64 subbands each.
Coarse up- and down-conversion is performed using a Polyphase Filter Bank (PFB).\cite{Harris2021, Vaidyanathan1990}
Each subband is sampled at 625 ksps and has a center frequency at integer multiples of 312.5 kHz.
This is a ``2× oversampled'' PFB, allowing complete capture of the input spectrum without subband aliasing.\cite{Harris2021}

In a PFB system, a band-defining filter is used to determine the spectral characteristics of each subband.
Typically, this band-defining filter is a compromise between several factors:

\begin{itemize}
	\item Passband ripple and stopband attenuation;
	\item The need to minimize aliasing without creating gaps in spectral coverage; and
	\item Computational complexity (or, equivalently, latency).
\end{itemize}

In our case, the signal path is enclosed within a feedback loop and latency becomes a driving design consideration to ensure stability with adequate loop gain.\cite{Smecher2022c}
The Dolph-Chebyshev window provides an optimally short window (and hence, low latency) subject to bandwidth and stopband attenuation constraints.\cite{Harris1978}
In our case, stopband attenuation is 100 dB (to match the dynamic range of a 16-bit input) and the bandwidth has been selected to give full spectral coverage without aliasing.
The resulting filter's length prior to polyphase decomposition is 256 taps, resulting in a latency (group delay) of about $6.4\si{\micro \second}$ for the window function.
(The full latency of the coarse downconverter must also include processing delays associated with the FFT, and the full latency of the system must also include the baseband processing and upconversion stages.)

\begin{figure}
	\centering
		\includegraphics{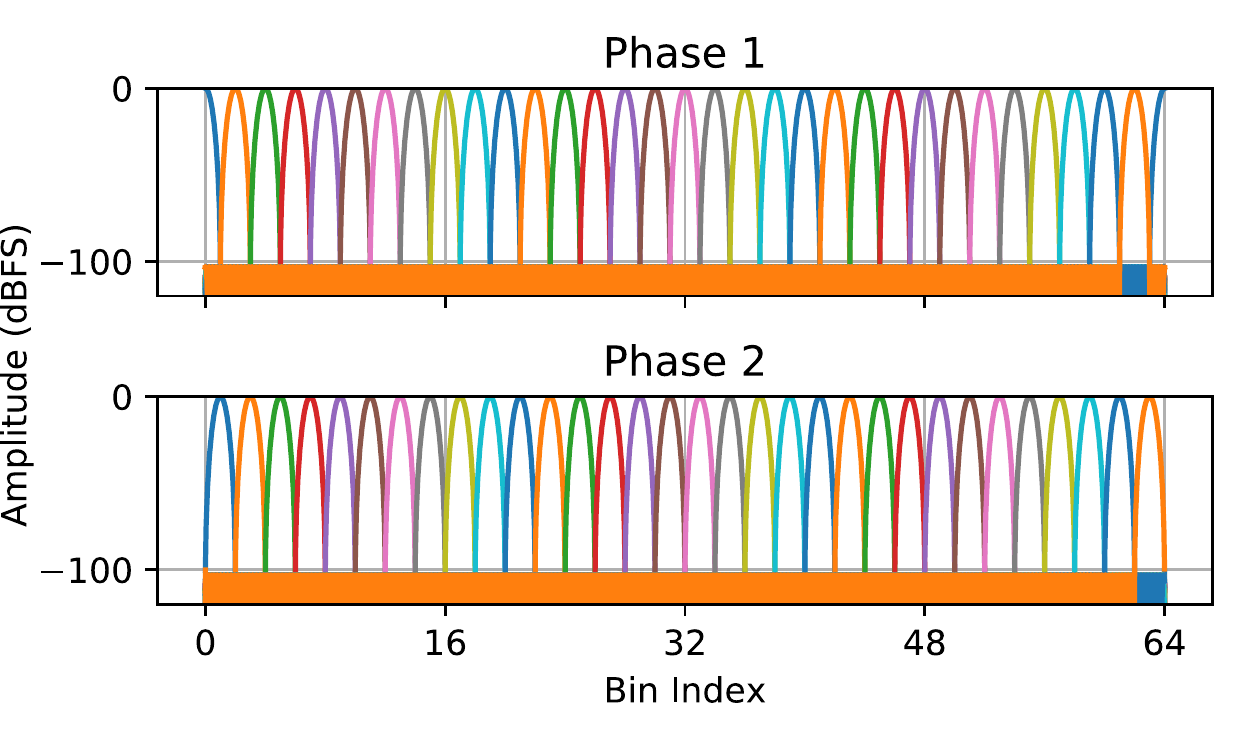}
		\caption{
			All 64 PFB subbands, separated into 2 non-overlapping groups (``phases'') for clarity.
			A bolometer at an arbitrary frequency will be fully attenuated in all but 2 of these subbands (one from the top group; one in the bottom group.)
			The subband with lower attenuation at the bolometer frequency is used to downconvert the bolometer to its baseband representation.
		}
		\label{fig:bin-band-shapes}
	%
\end{figure}

In exchange for bandwidth and latency optimality, however, the Chebyshev window exhibits significant gain variation across its passband (``droop'').
The band shapes associated with these subbands are shown in \Cref{fig:bin-band-shapes}.
Associated with each of these subbands is a well-defined stopband (the spectral region below -100 dB).
The frequency axis is normalized to subband count (i.e. an x-axis value of 64 corresponds to 20 MHz); the one-sided width of each passband (within which ripple is bounded) is 78.125 kHz.
In these subband units, the Nyquist frequency for each subband is 1 x-axis unit away from its band centre.
Compensation for this passband droop is performed alongside fine up-/down-conversion (see \Cref{fine-downconversion}.)

Because subbands overlap (\Cref{fig:bin-band-shapes}), an arbitrary point in the input spectrum is captured by 2 PFB subbands and fully attenuated by the remaining 62.
Of these two subbands, we always select the one with higher gain (lower attenuation) when performing fine downconversion between the subband centre and baseband.


\subsection{Fine Up-/Down-conversion}
\label{fine-downconversion}

The \acrshort{pfb} converts a 20 MSPS datastream into a sequence of 64 subband datastreams.
Each of the 128 baseband channels is formed from these 64 subband streams as follows.
First, each channel selects a sample from the \acrshort{pfb} subband nearest its desired frequency.
This sample is retrieved from a subband buffer that separates the timing of the PFB algorithm from the timing of the baseband signal path.
Second, a \acrshort{dds} forms a sample of the complex sinusoid necessary to complete downconversion to baseband.
This \acrshort{dds} sample is rotated by a user-specified phase coefficient and scaled by the necessary amplitude to compensate for \acrshort{pfb} ``droop'' described above.
Finally, the rotated and scaled \acrshort{dds} sample is multiplied by the subband sample to produce a single baseband sample for a single channel.
This process occurs channel-by-channel, re-using the same DDS and multipliers for each of the 128 channels per SQUID.

The downconversion and scaling process occurs in reverse during upconversion, in which case each channel's baseband sample contributes to a single PFB subband and must be renormalized to compensate for the droop that occurs during the synthesis \acrshort{pfb}.

After fine downconversion, signal-path samples use a single, consistent 24-bit I/Q encoding.
This encoding provides the necessary SNR to capture the dynamic range of the signal path's 16-bit, 20 MSPS inputs.
(Decimation by 32 implies a signal amplitude growth of 2.5 bits relative to an underlying white-noise floor.)
A 24-bit signal path matches the \acrshort{fpga}'s \acrshort{dsp} resources and is an acceptable match for the \acrshort{fpga}'s \acrshort{bram} resources.

\subsection{Baseband Processing}

All 128 bolometer channels associated with each SQUID channel are processed using the same resources on the \acrshort{fpga} using Time-Division Multiplexing (TDM).
Baseband processing consists of two identical feedback-loop controllers (``carrier'' and ``nuller''), each consisting of:

\begin{itemize}
	\item
		An 18-bit complex gain, which is used in \acrshort{dan} tuning to provide a digital loop gain\cite{Smecher2022c};
	\item
		An accumulator which can be enabled to provide an integrating feedback (e.g. \acrshort{dan}), or disabled for proportional-mode control;
	\item
		A programmable saturation stage, to restrict the amount of dynamic range that may be used by a single channel; and
	\item 
		A 24-bit complex offset, which can be used to provide static biasing when the rest of the loop controller is disabled (e.g. when configuring synthesizers in a feedback loop.)
\end{itemize}

Each parameter (gain, integrator enables, offset amplitudes) are programmed on a per-channel basis, allowing the dynamic behaviour and even the biasing mode of each bolometer to be separately controlled.

The baseband processor described here is a superset of the \acrshort{dan} controller described in Ref.~\citenum{deHaan2012} and is intended to enable other forms of feedback.
Accompanying the two feedback controllers in the baseband signal path are several other design elements:

\begin{itemize}
	\item
		Multiplexers that determine the origin of sampled timestreams for the carrier feedback controller and readout, and
	\item
		Ancillary samplers (not shown in \Cref{fig:dan-path}) used to gather short segments of data at various data rates (625 ksps, 20 MSPS) throughout the system.
		These capture buffers are not part of nominal operation and are used largely as diagnostic features.
\end{itemize}

\subsection{Readout}

The readout module consists of several chained signal-processing blocks:

\begin{itemize}
\item
  CIC1, a 3-stage CIC filter\cite{Hogenauer1981} with a decimation rate of 64, followed by
\item
  CIC2, a 6-stage CIC filter with a decimation rate of 64, followed by
\item
	A packetizer that combines science data and timestamps, forming packets which are sent to the \acrshort{dpu}.
\end{itemize}

Data generated by the packetizer are multiplexed with other data sources and transmitted to the \acrshort{dpu} as described in \Cref{sec:spacecraft-io}.

\subsubsection{CIC1}

\begin{figure}
	\centering
	\hfill
	\begin{subfigure}[b]{0.5\textwidth}
		\includegraphics[width=0.8\textwidth]{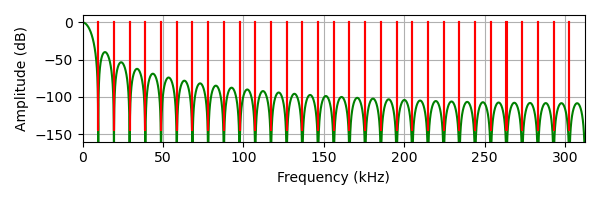}
		\caption{Wide-scale spectral response (input referred) of the CIC1 Design}
		\label{fig:cic1-wide-scale}
	\end{subfigure}
	\hfill
	\begin{subfigure}[b]{0.4\textwidth}
		\includegraphics[width=\textwidth]{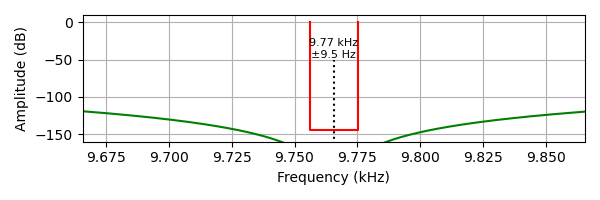}
		\caption{Narrow-scale CIC1 spectrum showing rejection of the first (worst) alias}
		\label{fig:cic1-zoom-first-alias}
	\end{subfigure}
	\hfill
	\vspace{0.2in}
	\caption{Spectral Characteristics of the first CIC1 Decimator.}
\end{figure}

The first \acrshort{cic} filter (CIC1) consists of a 3-stage decimate-by-64 \acrshort{cic} filter\cite{Hogenauer1981}.
All 128 bolometer channels are time-multiplexed onto the same \acrshort{cic} filter, producing an implementation that is compact and efficient.
The spectral response of CIC1 (over the entire input spectral range, from 0 Hz to 312.5 kHz, where 0 Hz is the baseband of each bolometer channel) is shown in \Cref{fig:cic1-wide-scale}.
Only the passband (near DC) and aliases (marked red) are relevant for the \acrshort{cic}'s performance.
The \acrshort{cic}'s actual performance -- the baseband (DC) droop and rejection of the worst (first) alias region -- are shown in detail in \Cref{fig:cic1-zoom-first-alias}.

The bolometer bandwidth for TES bolometer missions like LiteBIRD is anticipated to be less than approximately 10 Hz.
Hence, the alias rejection performance shown in \Cref{fig:cic1-zoom-first-alias} (greater than 144 dBfs) is better than the noise floor of the instrument.

\acrshort{cic} filters do not have flat passbands, and require amplitude compensation where passband flatness is desirable.\cite{Hogenauer1981}
However, because CIC1 is followed by a second CIC decimator, the passband droop created by CIC1 may be neglected.

\subsubsection{CIC2}

\begin{figure}
	\centering
	\hfill
	\begin{subfigure}[b]{0.5\textwidth}
		\includegraphics[width=0.8\textwidth]{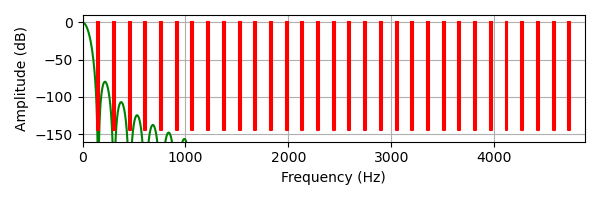}
		\caption{Wide-scale spectral response (input referred) of the CIC2 Design}
		\label{fig:cic2-wide-scale}
	\end{subfigure}
	\hfill
	\begin{subfigure}[b]{0.4\textwidth}
		\includegraphics[width=\textwidth]{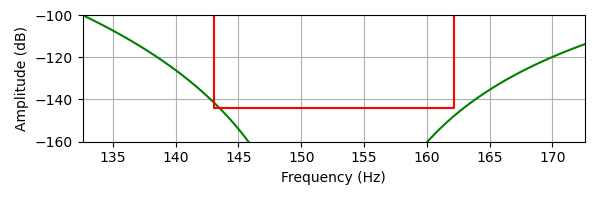}
		\caption{Narrow-scale CIC2 spectrum showing rejection of the first (worst) alias}
		\label{fig:cic2-zoom-first-alias}
	\end{subfigure}
	\hfill
	\vspace{0.2in}
	\caption{Spectral Characteristics of the CIC2 Decimator.}
\end{figure}

The second \acrshort{cic} filter (CIC2) is a 6-stage decimate-by-64 filter.
Unlike ground-based dfmux firmware, this filter has a fixed decimation rate.
The spectral response of CIC2 (over the entire input spectral range, from 0 Hz to 4.9 kHz, where 0 Hz is the baseband of each bolometer channel) is shown in \Cref{fig:cic2-wide-scale}.
The baseband (DC) droop and rejection of the worst (first) alias region are shown in detail in \Cref{fig:cic2-zoom-first-alias}.

We highlight two CIC2-specific differences between terrestrial dfmux deployments and this design:
\begin{itemize}
	\item
		In ground-based dfmux deployments, passband droop created by CIC2 is corrected by an \acrshort{fir} filter that decimates by a final factor of 2.\cite{Bender2014}
		Because it occurs at a very low sampling rate, this final signal-processing stage is resource-limited by the availability of memory and not computational fabric such as \acrshortpl{dsp}.
		For LiteBIRD, it is more convenient to offload this FIR to the \acrshort{dpu} where large amounts of RAM are already available.
	\item
		CIC2 is typically engineered in ground-based experiments to provide a variable decimation rate.
		(The compensating FIR described above is insensitive to CIC2's decimation rate.)
		For satellite-based experiments, readout bandwidth is relatively limited and the programmatic and technical complexity associated with supporting variable decimation rates is not justified.
\end{itemize}


\section{PERFORMANCE}
\label{performance}

\subsection{Resource Utilization}

Usage of \acrshort{fpga} resources is a driving design consideration for several reasons:
Firstly, sufficient resource margin must exist to manage programmatic risks associated with evolution of the firmware over time.
Secondly, resource usage is a proxy for power utilization and hence thermal stresses within the system.
(Power dissipation of the system is documented in Ref.~\citenum{Smecher2022a}.)
Thirdly, the design described here is portable to other parts from the same or different vendors, and the resource usage data presented here is correspondingly a good indicator of the design's scalability to other hardware platforms.

Resource utilization is summarized in \Cref{tab:resource-usage}.
The \acrshort{fpga} shown is an XQRKU060, which is Xilinx's current flagship space-grade part.
Although we focus on fabric resources, it is worth noting that the current design is I/O limited (it uses 80\% of the \acrshort{fpga}'s bonded I/Os.)
With either more I/Os, or with a more efficient use of the existing I/Os, the firmware could scale to accommodate additional SQUID modules.

\begin{table}
	\centering
	\begin{tabular}{l|l|l|l}
		\hline
		\textbf{Resource} & \textbf{Utilization} & \textbf{Available} & \textbf{\% Used} \\ \hline
		\acrshort{clb} \acrshortpl{lut} & 109,573 & 331,680 & 33\% \\
		\acrshort{clb} Registers & 181,860 & 663,360 & 27\% \\
		\acrshort{bram} & 502.5 & 1,080 & 47\% \\
		\acrshort{dsp} & 1,953 & 2,760 & 71\% \\ \hline
		\acrshort{iob} & 500 & 624 & 80\% \\ \hline
	\end{tabular}
	\caption{
		Top-level resource utilization of the firmware in an XQRKU060 \acrshort{fpga}.
		A detailed resource utilization breakout of the signal path is shown in \Cref{tab:resource-usage-detail}.}
	\label{tab:resource-usage}
\end{table}

\begin{table}
	\centering
	\begin{tabular}{l|r|r|r|r}
		\hline
		\textbf{Design Element} & \textbf{\acrshort{clb} \acrshortpl{lut}} & \textbf{\acrshort{clb} Registers} & \textbf{\acrshort{bram}} & \textbf{\acrshort{dsp}} \\ \hline
		\textbf{Single SQUID Module} & 6,696 & 11,049 & 31 & 122 \\
		~~\textbf{Ancillary Sampler} & 218 & 405 & 1.5 & 0 \\
		~~\textbf{Baseband Feedback} & 502 & 810 & 3.5 & 31 \\
		~~\textbf{Coarse Downconverter} & 1,953 & 3,650 & 3.5 & 31 \\
		~~\textbf{Coarse Upconverter} & 3,284 & 4,405 & 1 & 48 \\
		~~\textbf{Fine Up-/Down-converter} & 261 & 672 & 4.5 & 19 \\ \hline
		\textbf{All 16 SQUID Modules} & 107,136 & 176,784 & 496 & 1,952 \\ \hline
		\textbf{Entire Design} & 109,573 & 181,860 & 502.5 & 1,953 \\ \hline
		\textbf{Available Resources} & 331,680 & 663,360 & 1,080 & 2,760 \\ \hline
	\end{tabular}
	\caption{
		Detailed resource utilization for various firmware components.
		The signal path consumes over 97\% of all resource types in the system.
		Numbers in this table only sum approximately due to differences between how resources are attributed to blocks in the hierarchy.
	}
	\label{tab:resource-usage-detail}
\end{table}

The design's resource balance is conspicuous: it relies on specialized \acrshort{fpga} resources (\acrshortpl{bram}, \acrshortpl{dsp}) much more heavily than general-purpose resources (\acrshortpl{clb}).
We believe this balance to be a hallmark of power- and resource-efficient design.
This underuse of \acrshort{clb} resources should also have a positive effect on \acrshort{seu} susceptibility, since \acrshortpl{clb} take up the bulk of die area\cite{Lohrke2018} and are hence the largest target by volume for \acrshortpl{seu}.
(Relative \acrshort{seu} cross-sections for different resource types are available under NDA from the FPGA vendor.)

\begin{figure}
	\centering
	\includegraphics[width=0.65\textwidth, angle=-90]{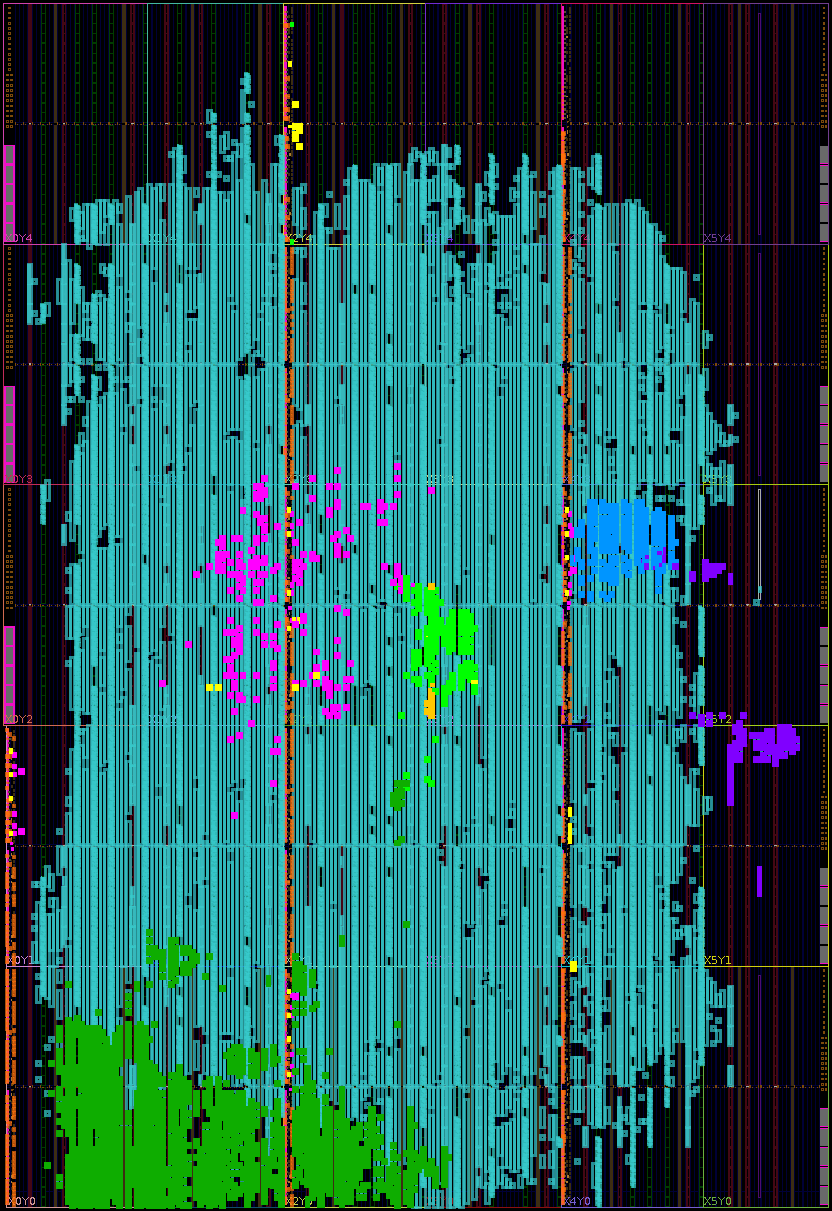}
	\caption{
		Top-level floorplan showing SPA firmware implemented on an XQRKU060 \acrshort{fpga}.
		This implementation supports 16 SQUID channels with 128 bolometers per SQUID, for a total of 2,048 bolometers.
		The design's ability to scale further is limited by I/O pins required to communicate with additional ADCs and DACs.
		The signal path for a single SQUID module (128 bolometers) is shown in dark green (top left).
		ADC/DAC interfaces are pink;
		clocking is bright yellow;
		peripheral interfaces (DA, SCA) are light blue;
		the SEU scrubber is purple;
		timestamp logic is dark yellow;
		the spacecraft control interface is bright green.
		The signal path (teal, plus the single SQUID channel highlighted in dark green) consumes the vast majority of the occupied \acrshort{fpga} fabric.
		Unused logic (70\% of regular fabric; 50\% of block RAMs; 30\% of DSPs) is black.
		This diagram shows the relative dominance of the signal path over all other logic.
	}
	\label{fig:top-level-floorplan}
\end{figure}

A top-level floorplan of the \acrshort{fpga} is shown in \Cref{fig:top-level-floorplan}.
(Contrast this floorplan to \Cref{fig:top-level-firmware}.)
A single SQUID module (128 bolometer channels) is highlighted in dark green; the other 15 SQUID channels are shown in teal.
All other firmware elements are coloured as described in the caption, and do not contribute meaningfully to the system's resource usage.

\begin{figure}
	\centering
	\includegraphics[width=0.85\textwidth]{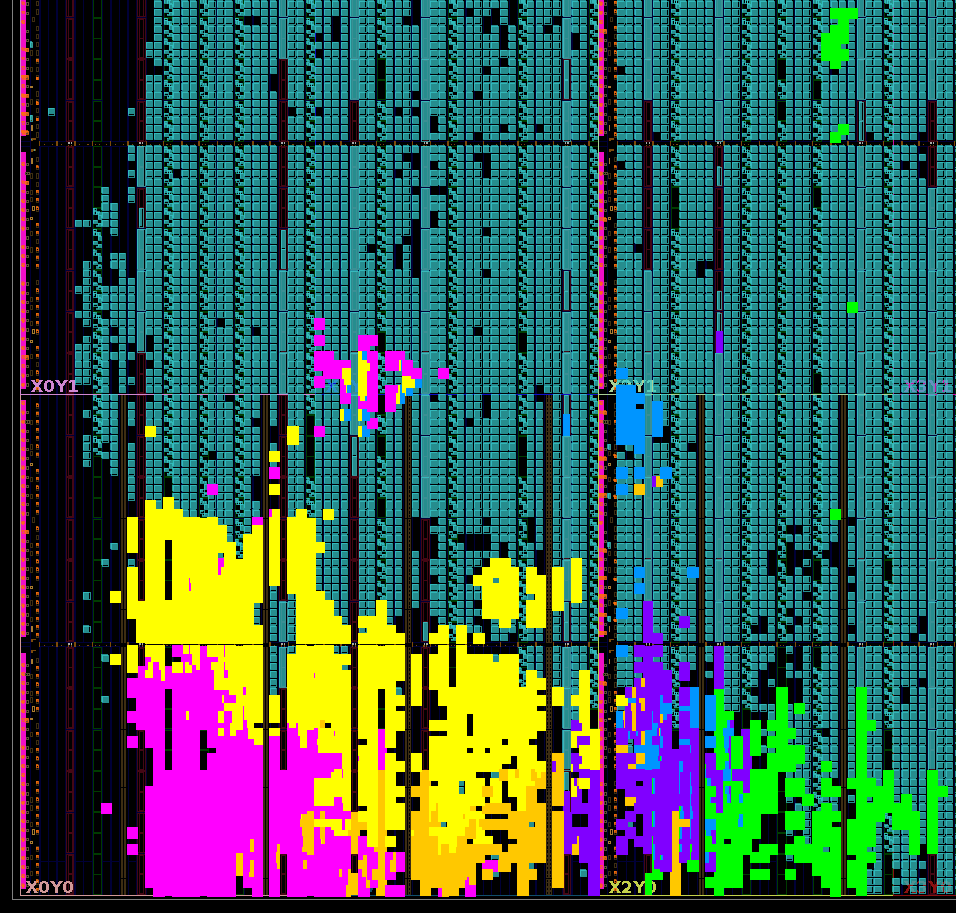}
	\caption{
		Detailed floorplan of a single SQUID module (the same module coloured in dark green in \Cref{fig:top-level-floorplan}; note rotation for formatting).
		Coarse downconversion is shown in pink;
		coarse upconversion is bright yellow;
		fine down/upconversion is dark yellow;
		feedback controllers are purple;
		readout (CIC1/CIC2/packetization) is green; and
		ancillary samplers are shown in blue.
		Resource usage is dominated by the \acrshortpl{fft} in the coarse up-/down-conversion blocks.
	}
	\label{fig:single-module-floorplan}
\end{figure}

\subsection{Power Consumption}

In this section, we provide preliminary power-consumption data on associated with the \acrshort{spu}'s digital electronics.
For these figures, analog electronics (including \acrshortpl{da}) are either absent or powered off; a characterization of system-level power dissipation is out of scope.
Measurements do not include \acrshort{pca} conversion losses.
The partial power-consumption figures here may be compared with results from an earlier hardware and firmware development platform in Ref.~\citenum{Bender2014}.
A complete summary of the SPU's power consumption (including \acrshort{pca} losses and analog electronics) may be found in Ref.~\citenum{Smecher2022a}.

The ``\acrshort{fpga} Unprogrammed'' state reflects the \acrshort{spa} while it is powered on, but before the \acrshort{fpga} has an active bitstream.
In this mode, all \acrshort{fpga} power rails are present but the \acrshort{fpga} is fully inactive.
This figure is a measurement of static power dissipation.

The ``\acrshort{fpga} Programmed, Channels Idle'' reflects the board when the \acrshort{fpga} is fully operational, but its signal path has not yet been configured (e.g. with channel frequencies and amplitudes).
All carrier, nuller, and demodulator channels are inactive.
The \acrshort{spa} responds to commands and produces readout packets populated with zeros.

The ``\acrshort{fpga} active, Channels Active'' test results below reflects the \acrshort{fpga} after all 16 SQUID channels have been configured for a multiplexing factor of 80.
This configuration uses arbitrary but representative frequencies and amplitudes.
The signal path is fully exercised, and power consumption is not artificially reduced.

\begin{table}
	\centering
	\begin{tabular}{l|l}
		\hline
		\textbf{State} & \textbf{Power Dissipation} \\ \hline
		\acrshort{fpga} Unprogrammed & 3.8 W \\
		\acrshort{fpga} Programmed, Channels Idle & 9.6 W \\
		\acrshort{fpga} Active, Channels Active & 12.0 W \\ \hline
	\end{tabular}
	\caption{
		Power consumption of a single \acrshort{spa}; digital supplies only.
		A complete treatment of \acrshort{spu} power dissipation may be found in Ref.~\citenum{Smecher2022a}.
	}
	\label{tab:power-consumption}
\end{table}

\subsection{Radiation Hardness}
\label{sec:conclusions-seu}

Payload electronics such as the \acrshort{spa} are not critical to spacecraft operations or survival.
As a result, it is possible to trade increased performance (e.g. design density, power/thermal efficiency) against susceptibility to tempary functional interruptions.
We expect the operation of the \acrshort{spu} to be disrupted by \acrfullpl{seu} in two ways:

\begin{itemize}
	\item
		Transient disruptions that produce a momentary glitch in device operation that is self-clearing, and
	\item
		Latching disruptions that produce a lasting glitch in device operation and require active intervention to repair.
\end{itemize}

Transient upsets can be filtered and discarded during analysis, and moreover, are expected to be generated by cosmic-ray impacts on the focal plane itself -- they cannot be fully mitigated by hardware or firmware design of the \acrshort{spu}.
Latching upsets require active intervention from either the \acrshort{dpu} or ground operations, and are hence more problematic.

Results from a preliminary investigation using \acrshort{seu} injection showed that

\begin{itemize}
	\item 95\% of \acrshortpl{seu} had no discernible effect;
	\item 4\% of \acrshortpl{seu} created a transient in readout; and
	\item 0.3\% of \acrshortpl{seu} created a lasting effect that would require active intervention from the \acrshort{dpu} or ground operations.
\end{itemize}

Dividing \acrshort{seu} effects into ``transient'' and ``latching'' is a phenomenological distinction, rather than a structural one.
This behaviour is a complex interaction between an \acrshortpl{seu} and the hardware, firmware, and software it acts on.
Design interventions throughout the system can change the relative occurrence of these effects and will co-evolve with the system.

\section{CONCLUSIONS}
\label{conclusions}

In this work, we described firmware for dfmux readout intended to operate in a satellite environment.
We provided a high-level structural overview of the system, and focused specifically on the design of the signal paths.
This firmware description is partnered with a hardware and system-level description provided in Ref.~\citenum{Smecher2022a}.

This firmware builds on a long history of ground-based dfmux deployments and has been optimized for space environments.
Optimizations include specific design interventions for radiation hardness and power efficiency, and the adoption of control and data interfaces suitable for spacecraft systems.
We provided preliminary performance metrics for resource consumption, \acrshort{seu} susceptibility, and power consumption.

Coupled with hardware, this firmware has been successfully operated with TES bolometers in a terrestrial test setting.
For missions such as LiteBIRD, the successful operation of this hardware and firmware platform represents an important milestone and is the natural starting point for integration activities to begin, both with spacecraft electronics (\acrshortpl{dpu}) and focal-plane electronics (\acrshortpl{scu}, \acrshortpl{squid}, \acrshortpl{tes}).



\acknowledgments
 
The authors gratefully acknowledge the support of the Canadian Space Agency (CSA), through Phase-0 (9F050-190058/001/MTB) and STDP (9F063-190285/003/MTB) contracts.

\bibliography{firmware}
\bibliographystyle{spiebib}

\end{document}